\documentclass[11pt,twoside]{article}

\usepackage{asp2006}
\usepackage{epsf}
\usepackage{psfig}
\usepackage{lscape}
\newcommand{\lsun}{\mbox{$L_\odot$}}
\newcommand{\msun}{\mbox{$M_\odot$}}
\newcommand{\hii}{H\mbox{\sc ~ii} }

\begin{document}
\title{Massive Infrared-Quiet Dense Cores: 
Unveiling the Initial Conditions of High-Mass Star Formation}

\author{F. Motte\altaffilmark{1}, S. Bontemps\altaffilmark{2, 1}, N. Schneider\altaffilmark{1}, P. Schilke\altaffilmark{3}, K. M. Menten\altaffilmark{3}} 
\altaffiltext{1}{Laboratoire AIM, CEA/DSM - CNRS - Univ. Paris Diderot, DAPNIA/SAp, CEA-Saclay, 91191 Gif-sur-Yvette Cedex, France}
\altaffiltext{2}{OASU/LAB, CNRS - Univ. Bordeaux 1, 2 rue de l'Observatoire, BP 89, 33270 Floirac, France}
\altaffiltext{3}{Max-Planck-Institut f\"ur Radioastronomie, Auf dem H\"ugel 69, 53121 Bonn, Germany}

%%%%%%%%%%%%%%% Abstract %%%%%%%%%%%%%%%%%%%%
\begin{abstract} 
As Pr. Th. Henning said at the conference, cold precursors of high-mass stars are now ``hot topics''. We here propose some observational criteria to identify massive infrared-quiet dense cores which can host the high-mass analogs of Class~0 protostars and pre-stellar condensations. We also show how far-infrared to millimeter imaging surveys of entire complexes forming OB stars are starting to unveil the initial conditions of high-mass star formation.
\end{abstract}

%%%%%%%%%%%%%%% 1. Introduction %%%%%%%%%%%%%%%%%%%%
\section{Introduction: First Discoveries of Cold Precursors of OB Stars}

For long, our understanding of high-mass star formation has been exclusively based on follow-up studies of bright sources found by \emph{IRAS} (see, e.g., reviews by Beuther et al. 2007; Zinnecker \& Yorke 2007). However, if the high-mass star formation process goes through cold, low-luminosity phases reminiscent of those of low-mass pre-stellar cores/condensations and Class~0s  (see Andr\'e et al. 2000), our knowledge has been biased against these earliest phases.

Recently, a few sources have been identified as good candidates for being cold precursors of high-mass stars, using mostly two methods. The first one uses high-density tracers (often submillimeter continuum) to map the surroundings of well-known \hii regions, H$_2$O or CH$_3$OH masers, or \emph{IRAS} sources. Some of these mappings have serendipitously revealed dense and massive cloud fragments which remain undetected at mid-infrared wavelengths (e.g., Motte et al. 2003; Hill et al. 2005; Klein et al. 2005; Beltr\'an et al. 2006). A second method is to search for sources seen in absorption against the diffuse mid-infrared background of square degrees images taken by \emph{ISO}, \emph{MSX}, and most recently \emph{Spitzer}. Indeed, these absorption features which are called infrared dark clouds (IRDCs) generally are the footprints of large-scale cloud structures that, in a few cases, harbor massive cold clumps (e.g., Carey et al. 2000; Hennebelle et al. 2001; Rathborne et al. 2006).

The sources identified in this way are definitively colder than the high-luminosity infrared sources taken to be the precursors of ultracompact (UC) \hii regions (like, e.g., HMPOs by Beuther et al. 2002). The evolutionary status (pre-stellar or protostellar) of these cold cloud fragments is unclear since very few of them have been surveyed for protostellar activity signatures other than infrared emission. Moreover, these new sources are generally located at large and inhomogeneous distances from the Sun and thus correspond  to cloud fragments with sizes ranging from 0.05~pc to 5~pc. As a consequence, only a handful of sources have been studied with enough spatial resolution, spectral energy distribution coverage, and follow-up studies to qualify as the high-mass equivalent of Class~0 protostars (Hunter et al. 1998; Molinari et al. 1998; Sandell 2000; Garay et al. 2002; Sandell \& Sievers 2004).

%%%%%%%%%%%%%%% 2. Criteria IR-quiet %%%%%%%%%%%%%%%%%%%%
\section{Observational Criteria for ``Massive Infrared-Quiet Dense Cores''}

To make progress, one needs to search, in a systematic and unbiased way, for high-mass analogs of pre-stellar cores/condensations, and Class~0 protostellar cores/protostars. Imaging surveys of entire complexes forming OB stars in the far-infrared or (sub)millimeter dust continuum are among the best tools to find candidate massive dense cores. Motte et al.  (2007) have made the first of such studies in the Cygnus~X molecular complex and identified 42 dense cores hosting either high-mass infrared-quiet protostars or high-luminosity infrared sources. From this experience, we establish three observational criteria to select, among cloud fragments, the best candidates for being  ``massive infrared-quiet dense cores/condensations'': (1) they should be small-scale so that the majority of their mass would be concerned by the star formation process in the near future; (2) they should be dense enough to permit the formation of one high-mass star rather than several low-mass stars; (3) they should have a bolometric luminosity smaller than that of a young stellar object (YSO) with a present stellar mass of $8~\msun$. The following subsections specify these criteria, illustrate them with our recent work on Cygnus~X, and interpret them  as physical constraints.

%%%%%%%%%%%%%%% 2.1 Small-scale %%%%%%%%%%%%%%%%%%%%
\subsection*{Small-Scale (0.01--0.1~pc) Cloud Fragments}

Table~1 summarizes a few reference studies that have searched for the precursors of UC \hii regions. For a meaningful comparison of the massive cloud fragments they identified, we choose to use the terminology recommended by Williams et al. (2000): ``clumps'' are  $\sim$1~pc cloud structures, ``dense cores'' have  $\sim$0.1~pc and ``condensations''  $\sim$0.01~pc sizes. 

For low-mass stars, the cloud fragments believed to be the direct precursors of single or binary stars are dense cores in distributed star-forming regions like Taurus and condensations in protoclusters such as $\rho$~Ophiuchi (see Table~1 and, e.g., Andr\'e et al. 2000). Therefore, if the physical process of high-mass star formation is not radically different from that for low-mass stars, we estimate that we must achieve the $0.01-0.1$~pc scale to get access to mass reservoirs used to form a single high-mass star. In that respect, most of the high-mass star formation studies have selected sources which have the size of clumps and thus probably harbor several protostars and pre-stellar condensations (see Table~1 and, e.g., Beuther et al. 2002; Rathborne et al. 2006). Motte et al.  (2007) have shown that focussing on the most compact cloud fragments located in the closest molecular cloud complexes is just enough to probe massive dense cores. High-resolution studies of these nearby dense cores are necessary to get access to high-mass starless condensations and individual high-mass protostars.

\begin{table}[htbp]
\caption[]{High-mass and low-mass cloud structures of a few reference studies}
\centering
\begin{tabular}{|l|ccc|cc|}
\hline
& HMPOs	& IRDCs & Cygnus~X	& Low-mass	& $\rho$~Oph\\
& clumps	& clumps	& dense cores	& dense cores & condens.\\
\hline
\hline
Diameter (pc)	& 0.5		& 0.5		& 0.13	& 0.08 	& 0.007\\
Mass (\msun)		& 290	& 150	 & 91 		& 4.7 	& 0.15 \\
$n_{\mbox{\tiny H$_2$}}$ (cm$^{-3}$)	& $8\times 10^3$ 	& $6\times 10^3$ 	 & $2\times 10^5$ & $3\times 10^4$	& $2\times 10^6$ \\
$d_{\mbox{\scriptsize Sun}}$ (kpc) & 0.3-14  & 1.8-7.1 & 1.7 & 0.14-0.44 & 0.14\\
References  & (1) & (2) & (3) & (4), (5) & (5)\\
\hline
\end{tabular}
References: (1) Beuther et al. (2002); (2) Rathborne et al. (2006); (3) Motte et al.  (2007); (4)  Ward-Thompson et al. (1999); (5) Motte et al. (1998).
\end{table}

%%%%%%%%%%%%%%% 2.2 High-density %%%%%%%%%%%%%%%%%%%%
\subsection*{Massive ($100~\msun$) and High-Density ($10^5$~cm$^{-3}$) Cores}
 
The large mass of a cloud structure is a necessary but not enough condition to ensure that a high-mass star is forming within it. The amount of gas encompassed by cloud structures harboring a high-mass protostar is also logically decreasing, from a few thousands to a few tens of solar mass, with its physical size shrinking from 1~pc to 0.01~pc. Therefore, it seems judicious to use both the mass and the volume-averaged density, $<n_{\mbox{\tiny H$_2$}}> \ =    M_{\mbox{\tiny smm}} \;/ \;(\frac{4}{3}\;\pi\times\mbox{\rm \it FWHM}^{\;3})$, of a cloud structure to estimate its capacity to form high-mass stars. 

As shown in Table~1, the most massive Cygnus~X dense cores ($\sim$0.1~pc and $\ge 40~\msun$, Motte et al. 2007) are, on average, 19 times more massive and 5 times denser than the pre-stellar dense cores forming low-mass stars (e.g., Motte et al. 1998; Ward-Thompson et al. 1999). They thus do not have any equivalent in the nearby star-forming regions and represent good candidate sites for forming high-mass stars. In contrast, many of the dense clumps found within infrared dark clouds should be neither massive enough, nor dense enough to form high-mass stars in the near future (see Table~1). In order to select good candidates for being precursors of high-mass stars, we estimate that a dense core should have a mass $\ge 50~\msun$ and a volume-averaged density $\ge 5\times 10^4$~cm$^{-3}$.

In that context, we have developed a source extraction technique aiming at identifying and extracting massive dense cores in (sub)millimeter continuum images of nearby complexes (Motte et al. 2003, 2007). In spirit, this method is equivalent to an eye search of local density peaks followed by flux measurements with an aperture optimized for each source. Automated, it separates dense cores from their environment by using a multi-resolution analysis of the cloud structures (MRE, Starck \& Murtagh 2006) and 2D-Gaussian fits of the compact cloud fragments (with the Gaussclumps program, Stutzki \&  G\"usten 1990). Setting a lower mass limit for the dense cores extracted by this technique selects the best potential sites of high-mass star formation. Indeed, in Cygnus~X, 42 dense cores have  {\it FWHM}$\ \sim 0.13$~pc, $M_{\mbox{\tiny smm}}\sim 91~\msun$, and $<n_{\mbox{\tiny H$_2$}}>\ 2\times 10^5$~cm$^{-3}$ and are probable precursors of high-mass stars (Motte et al. 2007).

%%%%%%%%%%%%%%% 2.3 IR-quiet %%%%%%%%%%%%%%%%%%%%
\subsection*{Massive Infrared-Quiet ($<$$10^3~\lsun$) Dense Cores}

When searching for the signature of embedded stellar embryos, the cross-cor\-relation of compact (sub)millimeter sources with infrared catalogs is crucial. Among the best-suited databases, the \emph{MSX}~C6 catalog provides fluxes at 8, 12, 15, and $21~\mu$m for point sources (with a $20\arcsec$ resolution, Egan et al. 2003) and the \emph{Spitzer}/MIPSGAL legacy project will soon publish point source catalogs at 24 and $70~\mu$m (with higher-resolution and higher-sensitivity, Benjamin et al. 2003). 
\emph{MSX} maps and catalogs have been extensively used to recognize bright infrared sources and distinguish them from colder cloud fragments (see, e.g., Fig.~1). However, a more quantitative criterium is necessary since a (sub)millimeter source qualified as ``cold'' on the basis of its non-detection by \emph{MSX} is surely weaker than the classical infrared-bright sources but could be, later on, detected by \emph{Spitzer}/MIPS.

\begin{figure}[htbp]
\vskip -1cm
\plottwo{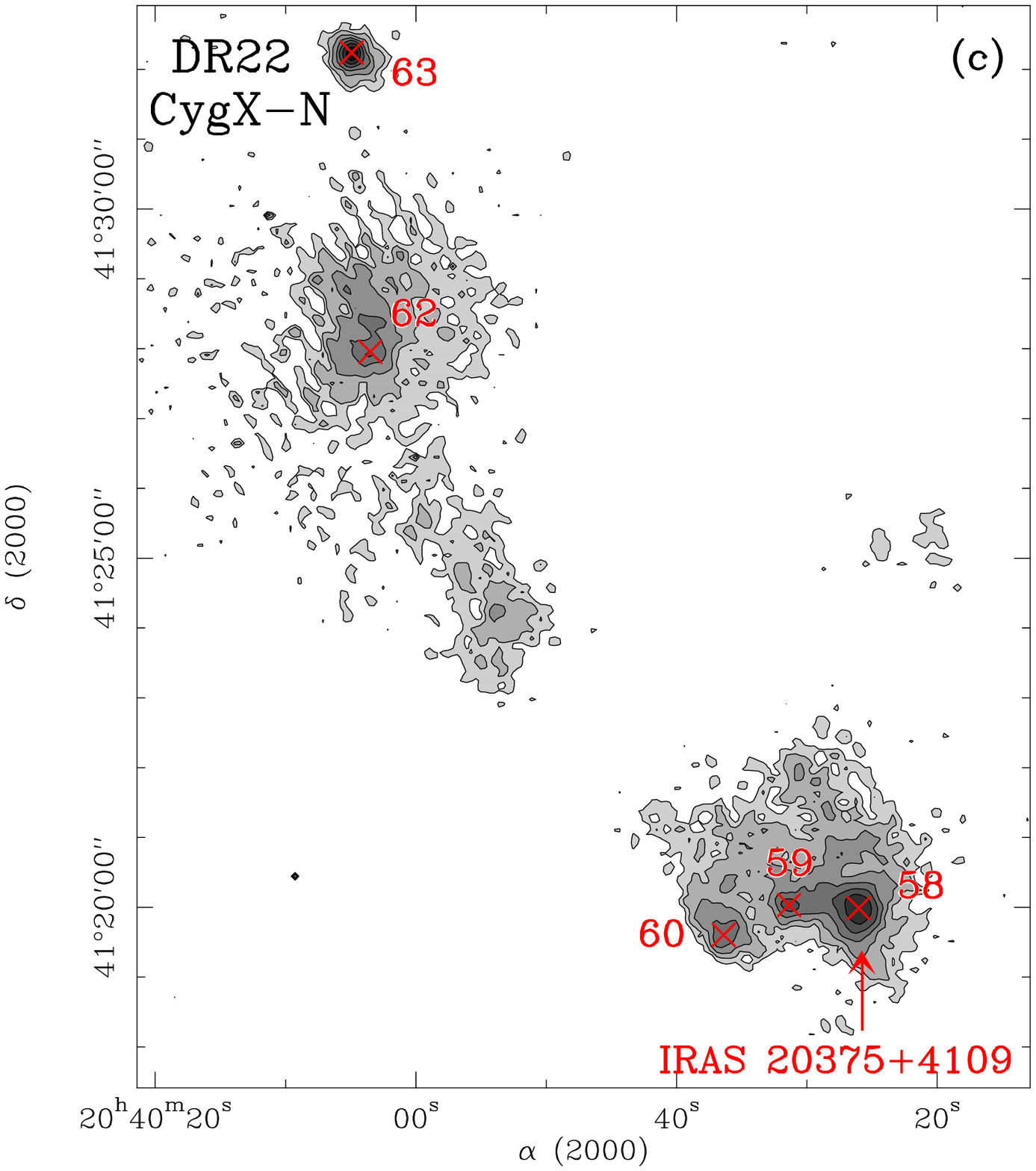}{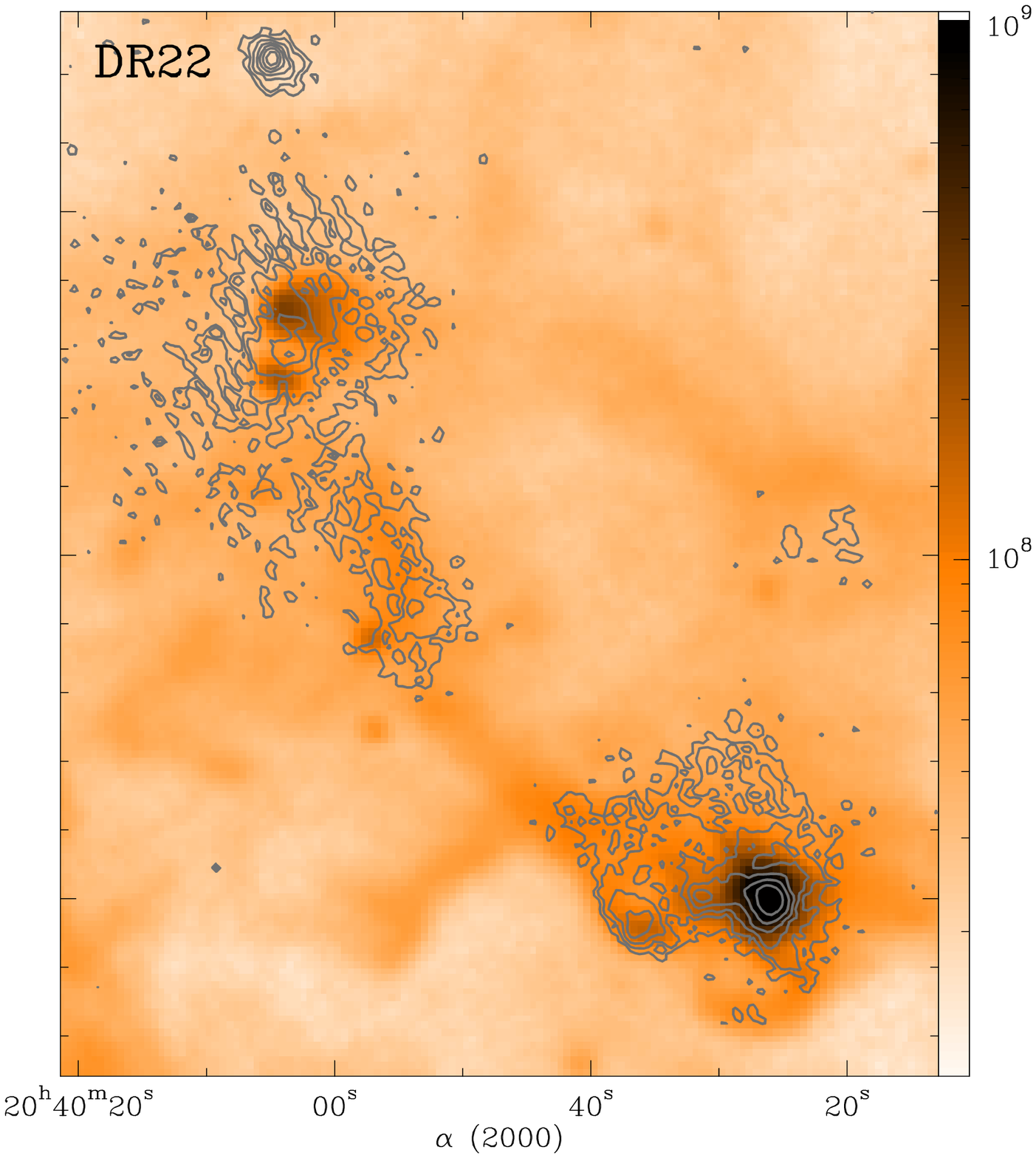}
\vskip -1cm
\caption[]{The DR22 region in Cygnus~X: {\bf left)} MAMBO-2 1.2~mm map in gray-scale and logarithmic contours, {\bf right)} 1.2~mm map in contours overlaid on the 8~$\mu$m image obtained by \emph{MSX}  (Jy sr$^{-1}$). Note that the massive dense cores \#62 and \#63 are seen in absorption against the mid-infrared background while \#58 is associated with a high-luminosity infrared source. By the kind permission of Motte et al. (2007) and EDP Sciences.}
\end{figure}

In the empirical evolutionary sequence of low-mass stars, Class~0 sources are young protostars at the beginning of the main accretion phase ($M_{\mbox{\tiny smm}} > M_\star \sim 1~\msun$) while Class~I sources are more evolved protostars which have already accumulated the majority of their final stellar mass ($M_{\mbox{\tiny smm}} < M_\star$). Applying this philosophy to high-mass stars forming in Cygnus~X, Motte et al.  (2007) propose to separate high-mass protostars that have already accreted more than $8~\msun$ from lighter and possibly younger protostars. In the context of massive dense cores harboring one high-mass plus a few low-mass protostars, a dense core qualifies as ``high-luminosity'' ($\ge$$10^3~\lsun$) if it can account for at least one B3 star on the zero-age main sequence. Conversely, a massive dense core is ``infrared-quiet'' if its bolometric luminosity is smaller than $10^3~\lsun$, meaning that the most massive of its protostellar embryos is $<$$8~\msun$. Since the bolometric luminosity of the massive dense cores extracted from (sub)millimeter images is usually not well constrained, Motte et al. have converted the $10^3~\lsun$ luminosity limit into a limit on their \emph{MSX} $21~\mu$m flux. To do so, they have assumed that the bolometric luminosity of these embedded sources is dominated by their mid- to far-infrared luminosity (see Motte et al. 2007 for details) and that their average \emph{IRAS} colors are defined by Wood \& Churchwell (1989). Finally, a massive infrared-quiet dense core should have a $21~\mu$m \emph{MSX} flux smaller than $\sim$10~Jy$\, \times\, (d_{\mbox{\scriptsize Sun}}/1.7~\mbox{kpc})^{-2} \, \times\, (L_{\mbox{\scriptsize bol}}/ 10^3~\lsun)$. 

 With such a luminosity or infrared flux criterium, 17 of the 42 massive dense cores of Cygnus~X qualify as infrared-quiet (see Fig.~2, Motte et al. 2007). Three of them are weak $21~\mu$m \emph{MSX} sources and we expect many more infrared-quiet dense cores (if not all of them, see Sect.~3) to be detected by \emph{Spitzer}/MIPS.

%%%%%%%%%%%%%%% FIG 7 %%%%%%%%%%%%%%
\begin{figure}[htbp]
\plotfiddle{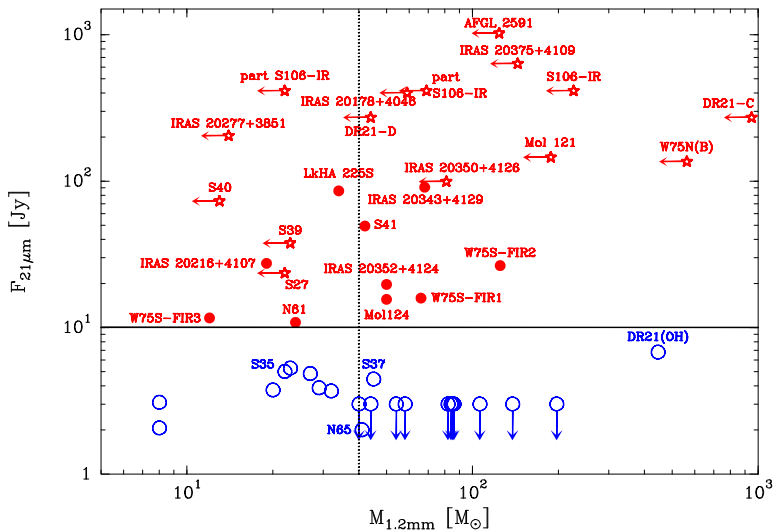}{6.9cm}{-90}{42}{42}{-180}{230}
\caption[]{Separating, in Cygnus~X, the high-luminosity infrared sources from massive infrared-quiet dense cores using their $21~\mu$m flux (limit set to 10~Jy). The high-luminosity infrared sources are either embedded UC\hii regions or \emph{IRAS} protostellar cores. The high-luminosity dense cores are not definitively more massive than massive infrared-quiet dense cores (blue open circles). By the kind permission of Motte et al. (2007) and EDP Sciences.}
\end{figure}

%%%%%%%%%%%%%%% 3. Nature %%%%%%%%%%%%%%%%%%%%
\section{The Nature of Massive Infrared-Quiet Dense Cores}

Massive infrared-quiet dense cores identified with the three criteria given in Sect.~2 could be either starless cores or protostellar cores in their earliest phase of evolution. Searching for protostellar activity signatures such as outflows, hot cores, or masers is necessary to finally determine their evolutionary status.

\begin{figure*}[htbp]
\centering
\plotone{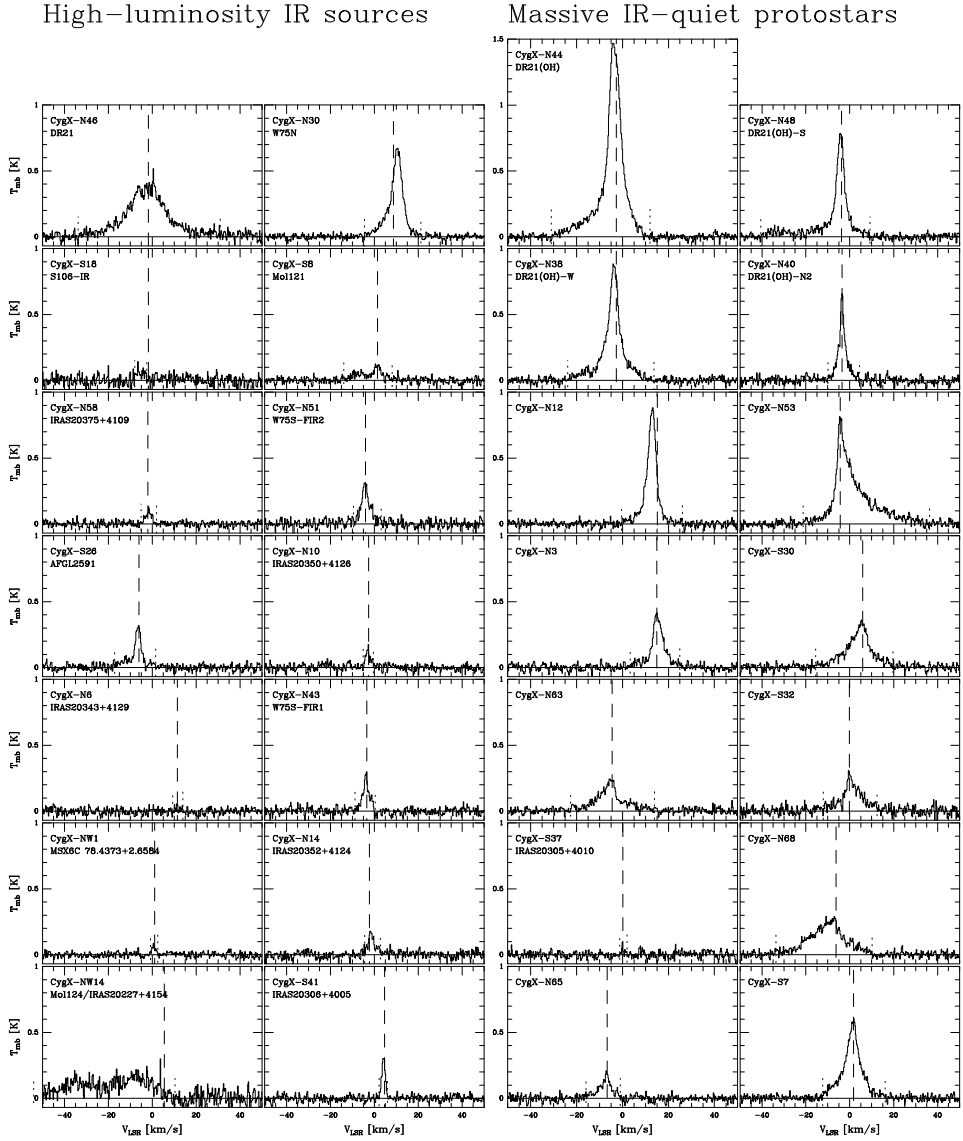}
\caption{SiO(2-1) lines observed toward the 28 most massive dense cores of Cygnus~X, and thus the best candidate sites to form high-mass stars: 14 high-luminosity infrared sources on the left and 14 massive infrared-quiet protostellar cores on the right, both ordered by decreasing $M_{\mbox{\tiny smm}}$ from top to bottom. Note that the SiO lines detected for massive infrared-quiet dense cores  are bright but  relatively simple with a main component and broad wings. By the kind permission of Motte et al. (2007) and EDP Sciences.}
\end{figure*}

Motte et al.  (2007) have surveyed the most massive dense cores of Cygnus~X in SiO(2-1) to search for shocked gas in molecular outflows. As shown in Fig.~3, the SiO lines detected for massive infrared-quiet dense cores in Cygnus~X are several times brighter than those observed for the high-luminosity sources. The association of high-velocity SiO emission with all massive infrared-quiet dense cores provides persuasive evidence that stars are already forming in each of them. Moreover, their SiO lines are, on average, 3 to 4 times brighter that the brightest SiO outflows of the nearby, low-mass protostars, which have a probability of occurence of only $6\%$. Therefore, the brightness and shape of the SiO lines suggest that these high-density cores may host single high-mass infrared-quiet protostars.

In summary, this first unbiased survey of the massive young stellar objects in Cygnus~X shows that high-mass infrared-quiet protostars do exist in large numbers, while the high-mass analogs of pre-stellar dense cores ($\sim$0.1~pc, $\sim$$10^5$~cm$^{-3}$) have not yet been convincingly identified.

%%%%%%%%%%%%%%% 4. Initial conditions %%%%%%%%%%%%%%%%%%%%
\section{Constraining the Initial Conditions of High-Mass Star Formation}

Our group has embarqued a multi-tracer imaging survey of Cygnus~X, one of the most active star-forming complexes at less than 3~kpc. According to Schneider et al. (2006), this massive ($4\times 10^6~\msun$) molecular complex is tightly associated with several OB associations (the largest being Cyg~OB2) and is located at only 1.7~kpc from the Sun. Studying all its high-density clouds will allow to set first constraints on the initial conditions of the high-mass star formation process, relevant at least in the Cygnus~X region.

The millimeter imaging survey of the entire Cygnus~X complex has provided a statistically significant sample of massive dense cores harboring high-mass pre-stellar condensations, protostars, or UC H\mbox{\sc ~ii}s (Motte et al. 2007). Since it is derived from a single molecular complex and covers every embedded phase of  high-mass star formation, it gives the first statistical estimates of their lifetime (see Table~2). Estimated relatively to the known age and numbers of OB stars in Cyg~OB2, the lifetimes of high-mass protostars and pre-stellar cores in Cygnus~X are $\sim$3$\times 10^4$~yr and $<$$10^3$~yr. In rough agreement with their free-fall time estimates, these statistical lifetimes are one and two order(s) of magnitude smaller, respectively, than what is found in nearby, low-mass star-forming regions (Kirk et al. 2005; Kenyon \& Hartmann 1995). These results suggest that high-mass pre-stellar and protostellar cores are in a highly dynamic state, as expected in, e.g., a molecular cloud where turbulent processes dominate.

\begin{table*}[htbp]
\caption[]{Lifetime/age estimates (in years) of massive YSOs in Cygnus~X}
\centering
\begin{tabular}{|l|ccc|cc|}
\hline
&  Pre-stellar	& IR-quiet		& high-lum	& \hii 	& OB\\
&  cores		& protostars	& protostars		& regions	& stars\\
\hline
\hline
Statistical lifetime
	& $\le$$8 \times 10^2$ 	& $1\times 10^4$ 	& $2 \times 10^4$ & $6\times 10^5$	& $2 \times 10^6$ \\
Free-fall time
	& $\sim$$9\times 10^4$ 	& \multicolumn{2}{c|}{$8\times 10^4$} & & \\
Low-mass lifetimes
          & $2\times 10^5$	& $2\times 10^4$	& $2\times 10^5$ & & \\
\hline
\end{tabular}
\end{table*}

%%%%%%%%%%%%%%% 5. Conclusion %%%%%%%%%%%%%%%%%%%%
\section{Conclusion and Perspectives: High-Mass Class 0-like Protostars}

With the advent of wide-field bolometer arrays on ground-based telescopes and the \emph{Herschel} satellite, we are entering a very promissing era for the studies of the cold phases of star formation.  Large-scale imaging surveys like ATLASGAL (with LABOCA on APEX by Schuller, Menten et al.) and HOBYS (with \emph{Herschel} by Motte, Zavagno, Bontemps et al.) will provide unbiased census of massive infrared-quiet dense cores and start constraining the initial conditions of high-mass star formation. \emph{Herschel} will define the spectral energy distribution of massive dense cores, thus allowing to measure their submillimeter to bolometric luminosity ratio that can replace the luminosity criterium for massive infrared-quiet dense cores.

However, the above surveys lack the necessary spatial resolution to look for the high-mass analogues of pre-stellar condensations and Class~0 protostars, within the massive dense cores. For instance, the Cygnus~X dense cores still are 20 times larger and 10 times less dense than the low-mass condensations of $\rho$~Ophiuchi which are believed to be the direct progenitors of single stars (see Table~1). High-resolution (sub)millimeter mapping of massive dense cores are therefore necessary first with, e.g.,  the IRAM Plateau de Bure interferometer and soon after with ALMA (see, e.g., Bontemps et al. in prep.).

%\acknowledgements 

%%% THE BIBLIOGRAPHY

\end{document}